# Flash Melting Amorphous Ice

Nathan J. Mowry[†], Constantin R. Krüger[†], Gabriele Bongiovanni[†], Marcel Drabbels, and Ulrich J. Lorenz[*]

**Affiliation:** Ecole Polytechnique Fédérale de Lausanne (EPFL), Laboratory of Molecular Nanodynamics, CH-1015 Lausanne, Switzerland

† These authors contributed equally

* To whom correspondence should be addressed. Email: ulrich.lorenz@epfl.ch

**Abstract**

Water can be vitrified if it is cooled at rates exceeding $3 \cdot 10^5$ K/s. This makes it possible to outrun crystallization in so-called no man's land, a range of deeply supercooled temperatures where water crystallizes rapidly. One would naively assume that the process can simply be reversed by heating the resulting amorphous ice at a similar rate. We demonstrate that this is not the case. When amorphous ice samples are flash melted with a microsecond laser pulse, time-resolved electron diffraction reveals that the sample transiently crystallizes despite a heating rate of more than $5 \cdot 10^6$ K/s, demonstrating that the critical heating rate for outrunning crystallization is significantly higher than the critical cooling rate during vitrification. Moreover, we observe different crystallization kinetics for amorphous solid water (ASW) and hyperquenched glassy water (HGW), which suggests that the supercooled liquids formed during laser heating transiently retain distinct non-equilibrium structures that are associated with different nucleation rates. These experiments open up new avenues for elucidating the crystallization mechanism of water and studying its dynamics in no man's land. They also add important mechanistic details to the laser melting and revitrification process that is integral to the emerging field of microsecond time-resolved cryo-electron microscopy.

Water is a poor glass former.[1,2] In order to achieve vitrification, it has to be cooled at rates exceeding $3 \cdot 10^5$ K/s,[3] which is necessary to outrun crystallization in a range of deeply supercooled temperatures between 160–232 K, in so-called no man's land,[1] where water crystallizes within tens of microseconds.[4] In fact, it had long been thought impossible that aqueous solutions could be vitrified,[5] making the demonstration of successful vitrification a stunning achievement.[6,7] Arguably, one of the most important applications of this process is the preparation of samples for cryo-electron microscopy (cryo-EM),[8] which is now set to become the preferred method in structural biology.[9] Vitrification allows biological specimens to be preserved in a frozen-hydrated state, in which they can be imaged in the vacuum of an electron microscope, while the damage inflicted by the electron beam is mitigated.[10–12]

Fast crystallization in no man's land had long made it impossible to map out the structural evolution of water during the vitrification process.[4,13–17] Recently, we have used time-resolved electron diffraction to show that water evolves smoothly from a high- to a low-temperature structure between 260 K and 220 K.[4] Below 200 K, its diffraction pattern converges to that of HGW, a glassy form of water, with a glass transition temperature of 136 K.[3] Here, we study the reverse of the vitrification process by characterizing the structural evolution of different amorphous ices during impulsive laser melting. Laser heating of amorphous ice has frequently been used to prepare deeply supercooled water.[15,18–22] For example, such experiments have provided evidence of a liquid-liquid phase separation in no man's land that can be observed after laser heating of high and low density amorphous ice.[21,22] Recently, we have also introduced a microsecond time-resolved approach to cryo-EM that involves rapidly melting a cryo sample with a microsecond laser pulse.[10,11,23] Once the sample is liquid, a suitable stimulus is used to initiate dynamics of the embedded proteins. As they unfold, the heating laser is switched off, and the sample revitrifies within microseconds, trapping the proteins in their transient configurations, in which they are subsequently imaged. For example, we have used this approach to image the microsecond motions that the capsid of the CCMV virus undergoes as part of its infection mechanism.[23] Previous studies have raised the question whether cryo samples may partially crystallize during laser melting.[11,24] Here, we address this question by characterizing the melting process with time-resolved electron diffraction.

Experiments are performed with a transmission electron microscope that we have modified for time-resolved experiments (Supplementary Methods 1).[25,26] As illustrated in Fig. 1a, a 263 nm thick layer of ASW is deposited *in situ* onto a sheet of few-layer graphene that is supported by a holey gold film (2 µm holes) on a 600 mesh gold grid held at 101 K. We then use a 30 µs laser pulse (532 nm) to melt the sample in the center of a grid square and probe its structural evolution by capturing a diffraction pattern with an intense, high-brightness electron pulse of 2 µs duration (Fig. 1b).

Figure 1c shows a micrograph of a typical ASW sample, with the corresponding diffraction pattern in the inset and the positions of the diffraction maxima indicated with blue lines (Supplementary Methods 12). When we heat the sample with a 30 µs laser pulse to about 185 K, it devitrifies and forms stacking disordered ice[27] (Fig. 1d), while at higher temperatures, the crystallization of hexagonal ice is favored[15,28] (Fig. 1e, ~235 K, Supplementary Methods 11). At even higher laser power, the sample melts and remains liquid for the duration of the laser pulse. Once the laser is switched off, the sample cools rapidly as the heat is efficiently dissipated to the surroundings, which have remained at cryogenic temperature.[4,10] With a cooling rate of over $10^7$ K/s, the sample vitrifies and forms HGW (Fig. 1f, after heating to 281 K). Note that the positions of the diffraction maxima of ASW in Fig. 1c are slightly shifted with respect to HGW. Such shifts are usually associated with the formation of microporous ASW, which is typically obtained at lower temperatures than in our experiment.[29,30] Contrast variations in the micrograph of Fig. 1c suggest that even under our deposition conditions, the samples exhibit a small amount of nanoscale heterogeneity.

We characterize the structural evolution of the ASW sample during laser melting with time-resolved electron diffraction, with the diffraction pattern of the ASW sample before the laser pulse shown in Fig. 2b (black curve). For comparison, a simulation of the temperature evolution during this process is presented in Fig. 2a (black curve, Supplementary Methods 8). Under laser irradiation, the sample heats up rapidly, reaching a temperature of 229 K at 6 µs. The corresponding diffraction pattern reveals that the supercooled liquid has traversed most of no man's land without crystallization (Fig. 2c). In contrast, additional diffraction features begin to appear at 7 µs (black arrows) that rapidly grow more intense (Fig. 2e, 9 µs), indicating the formation of stacking disordered ice, and later, hexagonal ice (Fig. S11). Once the sample temperature reaches the melting point (16.8 µs in the simulation of Fig. 2a), the crystallites begin to melt, so that at long time delays, the diffraction pattern turns into that of stable water (Fig. 2f, 25–30 µs). We determine that at the end of the laser pulse, the sample reaches a temperature of 281 K (Supplementary Methods 7).

Our experiments reveal somewhat counterintuitively that crystallization occurs during melting, but is avoided during vitrification, even though the respective heating and cooling rates are similar (Fig. 2a). In fact, when crystallization sets in at about 6.8 µs (Supplementary Methods 6), the heating rate has not dropped below $5 \cdot 10^6$ K/s. Evidently, the critical heating rate for outrunning crystallization is significantly higher than the critical cooling rate of about $3 \cdot 10^5$ K/s.[3] This can be qualitatively understood by considering how the nucleation and growth rates[1,31] (Fig. 2g) evolve during the melting and vitrification process (green and purple curves in Fig. 2a). As the sample heats up, a large number of nuclei are formed in no man's land where the nucleation rate goes through a maximum. At higher temperatures, nucleation slows significantly, but the growth rate picks up, causing the already existing nuclei to grow rapidly, so that the sample crystallizes. In contrast, during the vitrification process, a large concentration of nuclei is only formed once the growth rate has already dropped. These nuclei will then grow only little before the sample vitrifies and its structure is arrested.

Surprisingly, crystallization occurs significantly earlier when we repeat the experiment with the HGW sample that is obtained after the laser pulse (blue curves in Fig. 2b-f). Diffraction features indicating crystallizing already appear at 6 µs (Fig. 2c, blue arrows), after which they grow rapidly (Fig. 2d). At the end of the laser pulse, the sample has fully melted, and its diffraction pattern is indistinguishable from that of the ASW sample (Fig. 2f). The difference in crystallization kinetics is clearly evident in the evolution of the first diffraction maximum. Figure 3a shows that its intensity initially decreases as the sample heats up, but rises sharply when crystallization sets in, which occurs at about 6.8 µs for ASW (black), but already at 5.8 µs for HGW (blue), 1 µs earlier (Supplementary Methods 6). This difference cannot be fully explained by the fact that the HGW sample is thinner due to evaporation during the first laser pulse and therefore heats up more rapidly. Simulations of the crystallization kinetics reveal that this effect can only account for a time delay of about 0.48 µs (Supplementary Methods 9). Indeed, when we laser melt the HGW sample a second time, it only crystallizes another 0.42 µs earlier (HGW$_2$, green data points in Fig. 3a). The remaining time difference in the onset of crystallization between the ASW and HGW samples of over 0.5 µs must therefore result from a difference in the crystallization kinetics.

The intensity of the first diffraction peak goes through a maximum once the melting point is reached and crystal growth ceases, which occurs at approximately 8.0 µs for $HGW_2$, 8.5 µs for HGW, and 9.5 µs for ASW (Fig. 3b). Note that this is earlier than predicted by the simulation in Fig. 2a, which neglects that the crystallization process releases heat, so that the sample warms up more rapidly. This allows us to estimate that roughly a third of the sample crystallizes during laser melting (Supplementary Methods 10). Based on the relative diffraction intensities, we calculate that the maximum crystalline fraction of the HGW and $HGW_2$ samples is about 1.5 times as large as for ASW, a reflection of the faster crystallization kinetics of HGW.

The position of the first diffraction maximum provides insights into the structural evolution of the supercooled liquid before it crystallizes (Fig. 3b). The first diffraction maximum of HGW ($HGW_2$) appears at 1.74 Å$^{-1}$, whereas that of ASW is shifted to 1.77 Å$^{-1}$ under our deposition conditions, as discussed above. During laser heating, the peak positions barely change until crystallization sets in, which is marked by a shift to lower momentum transfer, indicating the formation of stacking disordered ice, which features a strong reflection at 1.71 Å$^{-1}$ (Fig. 1d). This suggests that the supercooled liquids obtained by laser heating ASW and HGW initially retain distinct structures. Nucleation, which predominantly takes place at lower temperatures (~140–230 K, Fig. 2g), must therefore occur from non-equilibrium configurations of supercooled water in our experiment. We speculate that these non-equilibrium structures possess different nucleation rates and that these lead to the observed difference in the crystallization kinetics of ASW and HGW during laser melting. Indeed, simulations have previously shown that different locally favored structures of water dictate its crystallization behavior, with pentagonal rings acting as a source of frustration.[32,33] Based on simulations of the crystallization kinetics, we estimate that the nucleation rate of HGW in no man's land is 3.4·10$^{26}$ m$^{-3}$s$^{-1}$, while that of ASW is 6.6·10$^{25}$ m$^{-3}$s$^{-1}$, about five times lower (Supplementary Methods 9). Note that our simulations can also reproduce the experimental observations, if we assume a 1.7 times higher growth rate for HGW, but an identical nucleation rate. We therefore cannot exclude that the observed difference in crystallization kinetics is in part due to different growth rates of the two supercooled liquids. This is however unlikely to be the dominant contribution since the growth rate scales with the diffusivity,[34] which intuitively should only weakly depend on changes in local structure.

In conclusion, our experiments reveal that samples of ASW and HGW partially crystallize during rapid melting with microsecond laser pulses, despite a heating rate of more than $5 \cdot 10^6$ K/s. We infer that the critical heating rate for outrunning crystallization must be more than one order of magnitude higher than the critical cooling rate during vitrification of about $3 \cdot 10^5$ K/s.[3] This can be understood by considering that as the sample heats up, it first undergoes fast nucleation before rapid growth sets in, whereas the opposite occurs during the vitrification process.[1,34] One can estimate an upper limit for the critical heating rate of $10^{10}$ K/s, since crystallization can be outrun with 10 ns laser pulses.[18–20]

Our analysis of the different crystallization kinetics of ASW and HGW suggests that laser heating these amorphous ices produces different non-equilibrium configurations of deeply supercooled water that exhibit different nucleation kinetics. This opens up the intriguing perspective of investigating how changes in local water structure affect the crystallization trajectory as well as the preference for the formation of different polymorphs.[35] For example, one would expect that preparing an amorphous ice precursor with an increased concentration in pentagonal rings should slow crystallization, which requires these rings to open and is therefore associated with a high energetic barrier.[32] Such experiments promise to elucidate the crystallization mechanism of water, which has so far remained elusive.[1,33]

We have previously shown that near-atomic resolution reconstructions can be obtained from revitrified cryo samples and that the structure of the proteins is not altered by the melting and revitrification process.[24,36] The partial crystallization of the sample during the melting process therefore does not appear to negatively affect microsecond time-resolved cryo-EM experiments. If it did, it should always be possible to increase the heating rate and outrun crystallization by using a shaped laser pulse with an intense leading edge. Transient crystallization during melting may however contribute to improving the sample quality by partially reshuffling the angular distribution of the particles[24] and thus help to overcome issues with preferred orientation that plague many cryo-EM projects.[37]

**Acknowledgments:**

The authors would like to thank Dr. Pavel K. Olshin for his help with the heat transfer simulations as well as Dr. Jonathan M. Voss for his help with the preparation of Fig. 1.

**Funding:**

This work was supported by the ERC Starting Grant 759145 as well as by the Swiss National Science Foundation Grants PP00P2_163681 and 200020_207842.

**Author contributions:**

Conceptualization: UJL

Methodology: NJM, CRK, GB, UJL

Investigation: NJM, CRK, GB, UJL

Visualization: NJM, CRK, GB, UJL

Funding acquisition: UJL

Project administration: UJL

Supervision: UJL

Writing – original draft: NJM, CRK, GB, UJL

Writing – review & editing: NJM, CRK, GB, MD, UJL

**Competing interests:**

The authors declare that they have no competing interests.

**Data and materials availability:**

All data are available from the corresponding author upon request.

**References**

1. Gallo, P. *et al.* Water: A Tale of Two Liquids. *Chem. Rev.* **116**, 7463–7500 (2016).

2. Handle, P. H., Loerting, T. & Sciortino, F. Supercooled and glassy water: Metastable liquid(s), amorphous solid(s), and a no-man's land. *Proc. Natl. Acad. Sci.* **114**, 13336–13344 (2017).

3. Warkentin, M., Sethna, J. P. & Thorne, R. E. Critical Droplet Theory Explains the Glass Formability of Aqueous Solutions. *Phys. Rev. Lett.* **110**, 015703 (2013).

4. Krüger, C. R., Mowry, N. J., Bongiovanni, G., Drabbels, M. & Lorenz, U. J. Electron diffraction of deeply supercooled water in no man's land. *Nat. Commun.* **14**, 2812 (2023).

5. Dubochet, J. Cryo-EM—the first thirty years. *J. Microsc.* **245**, 221–224 (2012).

6. Dubochet, J. & McDowall, A. W. VITRIFICATION OF PURE WATER FOR ELECTRON MICROSCOPY. *J. Microsc.* **124**, 3–4 (1981).

7. Brüggeller, P. & Mayer, E. Complete vitrification in pure liquid water and dilute aqueous solutions. *Nature* **288**, 569–571 (1980).

8. Cheng, Y., Grigorieff, N., Penczek, P. A. & Walz, T. A Primer to Single-Particle Cryo-Electron Microscopy. *Cell* **161**, 438–449 (2015).

9. Hand, E. Cheap Shots. *Science* **367**, 354–358 (2020).

10. Voss, J. M., Harder, O. F., Olshin, P. K., Drabbels, M. & Lorenz, U. J. Rapid melting and revitrification as an approach to microsecond time-resolved cryo-electron microscopy. *Chem. Phys. Lett.* **778**, 138812 (2021).

11. Voss, J. M., Harder, O. F., Olshin, P. K., Drabbels, M. & Lorenz, U. J. Microsecond melting and revitrification of cryo samples. *Struct Dyn* **8**, 054302 (2021).

12. Glaeser, R. M. Specimen Behavior in the Electron Beam. in *Methods in Enzymology* vol. 579 19–50 (Elsevier, 2016).

13. Kimmel, G. A. *et al.* Homogeneous ice nucleation rates and crystallization kinetics in transiently-heated, supercooled water films from 188 K to 230 K. *J. Chem. Phys.* **150**, 204509 (2019).

14. Kringle, L., Thornley, W. A., Kay, B. D. & Kimmel, G. A. Structural relaxation and crystallization in supercooled water from 170 to 260 K. *Proc. Natl. Acad. Sci.* **118**, e2022884118 (2021).

15. Ladd-Parada, M. *et al.* Following the Crystallization of Amorphous Ice after Ultrafast Laser Heating. *J. Phys. Chem. B* **126**, 2299–2307 (2022).

16. Sellberg, J. A. *et al.* Ultrafast X-ray probing of water structure below the homogeneous ice nucleation temperature. *Nature* **510**, 381–384 (2014).

17. Laksmono, H. *et al.* Anomalous Behavior of the Homogeneous Ice Nucleation Rate in "No-Man's Land". *J. Phys. Chem. Lett.* **6**, 2826–2832 (2015).

18. Kringle, L., Thornley, W. A., Kay, B. D. & Kimmel, G. A. Reversible structural transformations in supercooled liquid water from 135 to 245 K. *Science* **369**, 1490–1492 (2020).

19. Xu, Y., Petrik, N. G., Smith, R. S., Kay, B. D. & Kimmel, G. A. Homogeneous Nucleation of Ice in Transiently-Heated, Supercooled Liquid Water Films. *J. Phys. Chem. Lett.* **8**, 5736–5743 (2017).

20. Xu, Y. *et al.* A nanosecond pulsed laser heating system for studying liquid and supercooled liquid films in ultrahigh vacuum. *J. Chem. Phys.* **144**, 164201 (2016).

21. Amann-Winkel, K. *et al.* Liquid-liquid phase separation in supercooled water from ultrafast heating of low-density amorphous ice. *Nat. Commun.* **14**, 442 (2023).

22. Kim, K. H. *et al.* Experimental observation of the liquid-liquid transition in bulk supercooled water under pressure. *Science* **370**, 978–982 (2020).

23. Harder, O. F., Barrass, S. V., Drabbels, M. & Lorenz, U. J. Fast viral dynamics revealed by microsecond time-resolved cryo-EM. *Nat. Commun.* **14**, 5649 (2023).

24. Bongiovanni, G., Harder, O. F., Voss, J. M., Drabbels, M. & Lorenz, U. J. Near-atomic resolution reconstructions from in situ revitrified cryo samples. *Acta Crystallogr. Sect. D* **79**, 473–478 (2023).

25. Bongiovanni, G., Olshin, P. K., Drabbels, M. & Lorenz, U. J. Intense microsecond electron pulses from a Schottky emitter. *Appl. Phys. Lett.* **116**, 234103 (2020).

26. Olshin, P. K., Bongiovanni, G., Drabbels, M. & Lorenz, U. J. Atomic-Resolution Imaging of Fast Nanoscale Dynamics with Bright Microsecond Electron Pulses. *Nano Lett.* **21**, 612–618 (2021).

27. Kuhs, W. F., Sippel, C., Falenty, A. & Hansen, T. C. Extent and relevance of stacking disorder in 'ice Ic'. *Proc. Natl. Acad. Sci.* **109**, 21259–21264 (2012).

28. Amann-Winkel, K. *et al.* Water's second glass transition. *Proc. Natl. Acad. Sci.* **110**, 17720–17725 (2013).

29. Hallbrucker, A., Mayer, E. & Johari, G. P. Glass-liquid transition and the enthalpy of devitrification of annealed vapor-deposited amorphous solid water: a comparison with hyperquenched glassy water. *J. Phys. Chem.* **93**, 4986–4990 (1989).

30. Stevenson, K. P., Kimmel, G. A., Dohnálek, Z., Smith, R. S. & Kay, B. D. Controlling the Morphology of Amorphous Solid Water. *Science* **283**, 1505–1507 (1999).

31. Debenedetti, P. G. *Metastable Liquids: Concepts and Principles*. (Princeton University Press, 1997).

32. Russo, J. & Tanaka, H. Understanding water's anomalies with locally favoured structures. *Nat. Commun.* **5**, 3556 (2014).

33. Russo, J., Romano, F. & Tanaka, H. New metastable form of ice and its role in the homogeneous crystallization of water. *Nat. Mater.* **13**, 733–739 (2014).

34. Xu, Y., Petrik, N. G., Smith, R. S., Kay, B. D. & Kimmel, G. A. *Growth rate of crystalline ice and the diffusivity of supercooled water from 126 to 262 K*. vol. 113 (Proc Natl Acad Sci USA, 2016).

35. Seidl, M., Amann-Winkel, K., Handle, P. H., Zifferer, G. & Loerting, T. From parallel to single crystallization kinetics in high-density amorphous ice. *Phys. Rev. B* **88**, 174105 (2013).

36. Bongiovanni, G., Harder, O. F., Drabbels, M. & Lorenz, U. J. Microsecond melting and revitrification of cryo samples with a correlative light-electron microscopy approach. *Front. Mol. Biosci.* **9**, 1044509 (2022).

37. Glaeser, R. M. How good can cryo-EM become? *Nat. Methods* **13**, 28–32 (2016).

**Sample geometry and experimental approach**

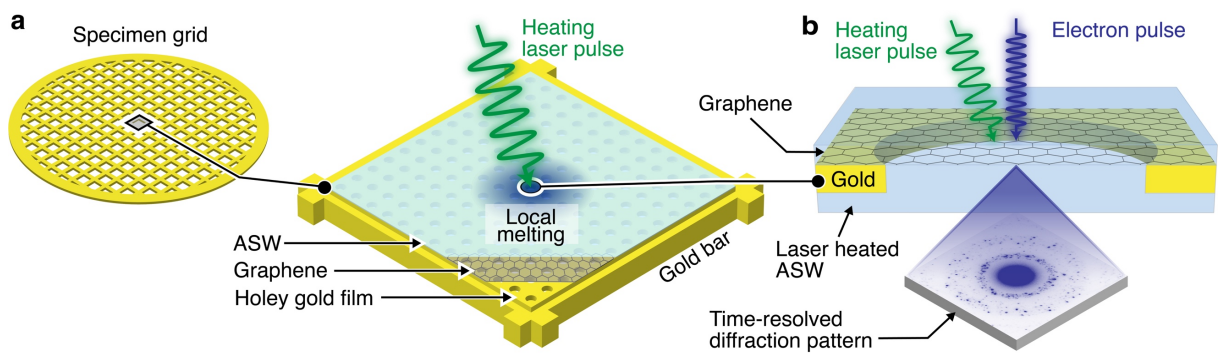

**Different amorphous and crystalline ices prepared through laser heating of ASW**

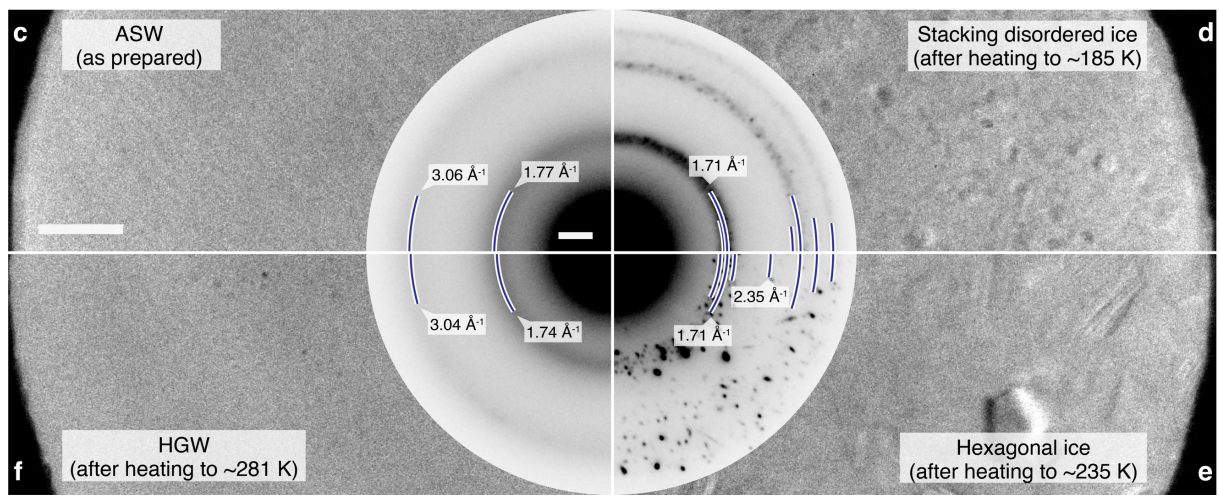

**Figure 1 | Illustration of the experimental approach as well as micrographs and diffraction patterns of different amorphous and crystalline ices created in our experiment. a** Illustration of the sample geometry. A gold mesh supports a holey gold film covered with multilayer graphene. A 263 nm thick layer of ASW is deposited (101 K sample temperature) and locally melted with a 30 µs laser pulse. **b** Diffraction patterns of the structural evolution of the ASW sample during melting process are captured with intense, 2 µs electron pulses (200 kV accelerating voltage). **c-f** Micrographs and diffraction patterns of different amorphous and crystalline ices that are created in our experiment. The ASW sample (**d**) crystallizes into stacking disordered ice when heated to a temperature of about 185 K with a 30 µs laser pulse (**e**), while hexagonal ice is mostly formed at about 235 K (**f**). When the sample is heated to about 281 K, it melts and vitrifies after the laser pulse to form HGW. The diffraction maxima are indicated with blue lines. Scale bars are 150 nm and 1 Å⁻¹.

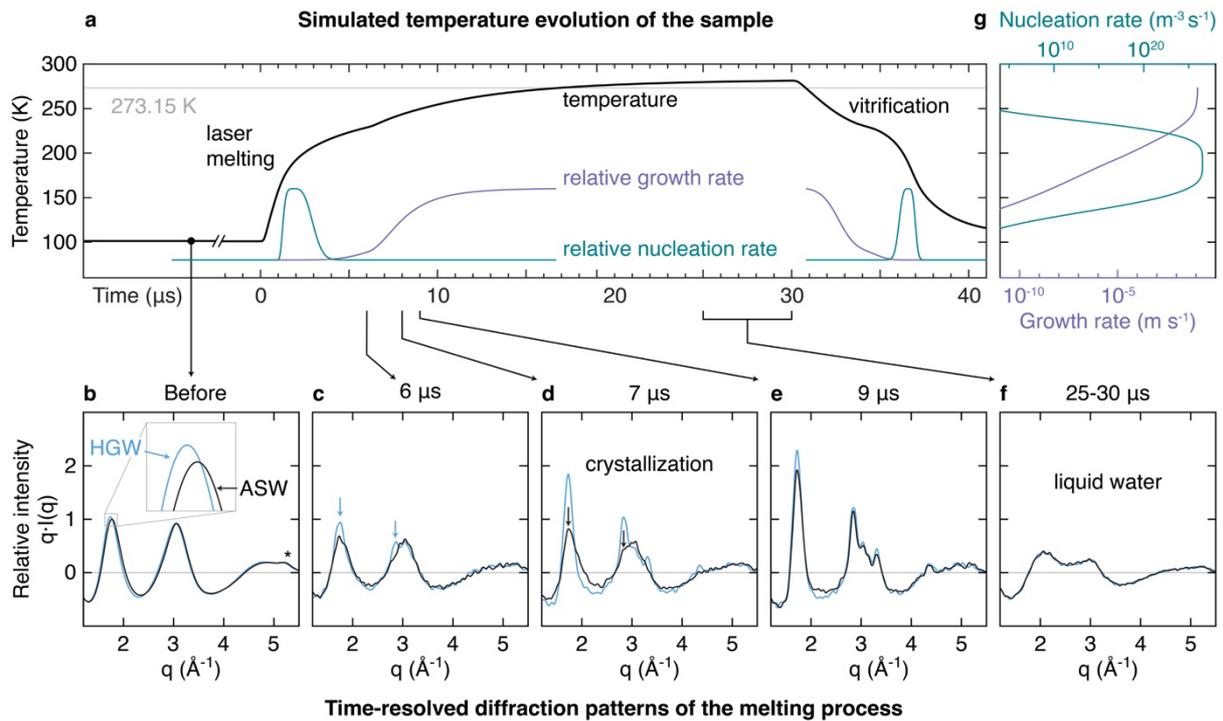

**Figure 2 | Time-resolved diffraction patterns capture the structural evolution of the sample during the laser melting process. a** Simulated temperature evolution of the sample (black). The evolution of the relative nucleation and growth rates that the sample experiences during melting and vitrification are shown in green and purple, respectively. **b-f** Time-resolved diffraction patterns of the melting process for ASW (black) and HGW (blue). Arrows indicate the appearance of crystalline features. The asterisk in **b** marks a diffraction feature arising from the graphene support. **g** Absolute nucleation and growth rates (green and purple, respectively) as a function of temperature.[1,31]

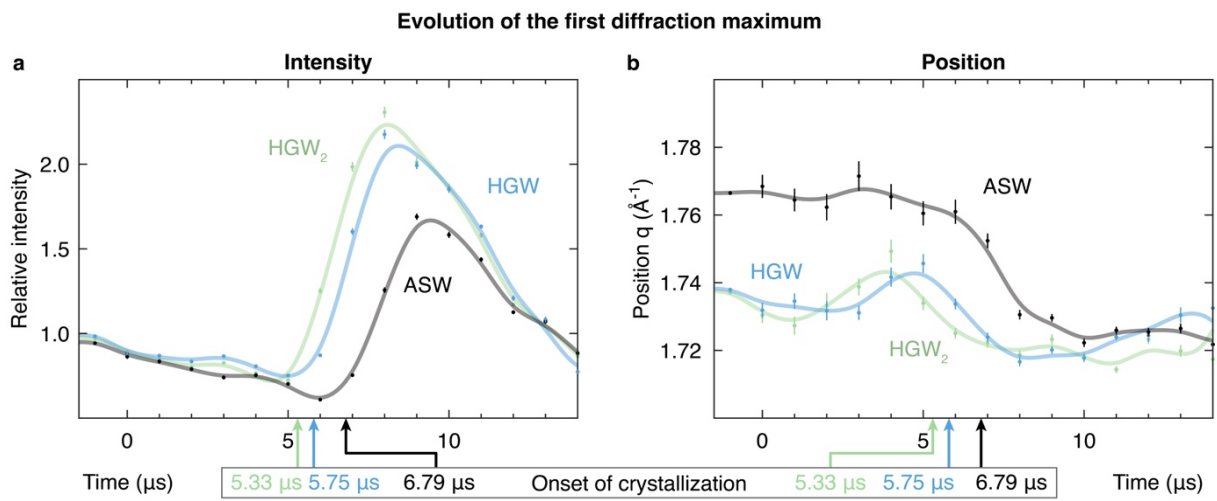

**Figure 3 | The temporal evolution of the first diffraction maximum reveals the crystallization and structural relaxation dynamics of the sample. a** Evolution of the diffraction intensity for ASW (black) and HGW (blue), as obtained after melting and vitrification of the ASW sample, as well as for HGW after melting and revitrifying a second time (HGW₂, green). **b** Evolution of the position of the diffraction maximum. The solid lines provide a guide to the eye and are derived from splines. Error bars indicate standard errors of the fit used to determine the intensities and positions.

**Supplementary Information**

**Flash Melting Amorphous Ice**

Nathan J. Mowry[†], Constantin R. Krüger[†], Gabriele Bongiovanni[†], Marcel Drabbels,

and Ulrich J. Lorenz[*]

**Affiliation:** Ecole Polytechnique Fédérale de Lausanne (EPFL), Laboratory of Molecular Nanodynamics, CH-1015 Lausanne, Switzerland

**This PDF file includes:**

[†] These authors contributed equally.

[*] To whom correspondence should be addressed. E-mail: ulrich.lorenz@epfl.ch

# 1 Instrumentation

Experiments were performed with a modified JEOL 2010F transmission electron microscope that we have previously described (Fig. S1).(*1, 2*) Amorphous solid water (ASW) is deposited onto a holey gold film, which is flash heated *in situ* by irradiation with a microsecond laser pulse (532 nm, 30 μs duration, typically 100 mW power). The microsecond laser pulse is obtained by modulating the output of a continuous laser with an acousto-optic modulator. The laser beam is directed at the sample with a mirror mounted above the upper pole piece of the objective lens, so that it strikes the sample at close to normal incidence. The laser beam is focused to a spot size of 38 μm FWHM in the sample plane, as determined from an image of the laser beam recorded with a CCD camera that is placed in a conjugate plane.

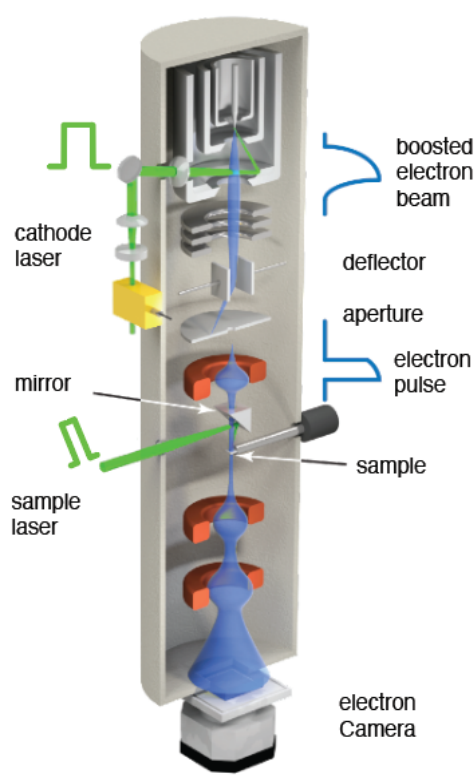

**Figure S1. Illustration of the time-resolved electron microscope.** Illustration of the time-resolved electron microscope used for preparing and observing the crystallization dynamics of amorphous solid water and hyperquenched glassy water.

The structural evolution of the sample during laser heating is then probed at a well-defined point in time by capturing a diffraction pattern with an intense, high-brightness electron pulse.(*1, 2*) We generate such electron pulses as previously described, by temporarily boosting the emission from the Schottky emitter of our microscope to near its limit. To this end, we briefly heat the emitter tip to extreme temperatures through irradiation with a microsecond laser pulse (532 nm, 1.5 W, 17 µm FWHM spot size at the tip, 100 µs pulse duration, as obtained by chopping the output of a continuous laser with an acousto-optic modulator), which causes the emission current to increase by up to 3.7 times. An electrostatic deflector is then used to slice a 2 µs electron pulse out of the boosted electron beam. As illustrated in Fig. 1a,b, time-resolved electron diffraction patterns are collected from within the central hole of a grid square of the holey gold specimen support (see Supporting Note 8 for a detailed description of the sample geometry), with the electron beam converged to a disk of about 1.5 µm diameter. The diffraction patterns are recorded with a TVIPS XF416 electron camera. The camera length is calibrated with the diffraction pattern of the polycrystalline holey gold film.(3)

## 2 Fabrication of sample supports

Sample supports are fabricated as previously described, following the process illustrated in Fig. S2.(*4*) A holey gold film is prepared by vapor depositing 50 nm of gold onto a holey carbon specimen grid (QUANTIFOIL N1-C15nCu20-01) which consists of a 12 nm holey carbon film (2 µm diameter holes with 1 µm spacing) on a 200 mesh copper grid (Fig. S2a,b). The copper mesh is then etched away by floating the grid on an ammonium persulfate solution, until only the holey thin film is left (Fig. S2c). The remaining etchant is removed in three washing steps, each consisting in transferring the thin film to a fresh bath of deionized water for approximately 10 min. The holey thin film is then transferred onto a 600 mesh gold grid (Plano, 13.5 µm bar width and 8.75 µm bar height) by submerging the grid into the water bath and using it to gently pick up the film (Fig. S2d). The assembly is placed onto a hot plate (50 °C) for up to 2 hours to evaporate any remaining water. In the final step, multilayer graphene is transferred onto the grid assembly. To this end, 6-8 layer graphene on copper foil (Graphene Supermarket) is floated on an ammonium persulfate solution until the copper has dissolved. The multilayer graphene film, which remains floating on the etchant, is then washed three times by transferring it to a fresh bath of distilled water for 10 min each. Finally, the graphene layer is transferred onto the holey gold film by gently picking it up with the specimen grid (Fig. S2e). Before use, the completed assembly is cleaned for 40 seconds in a hydrogen plasma (Pelco Easiglow, negative polarity, 20 mA current, 0.5 mbar).

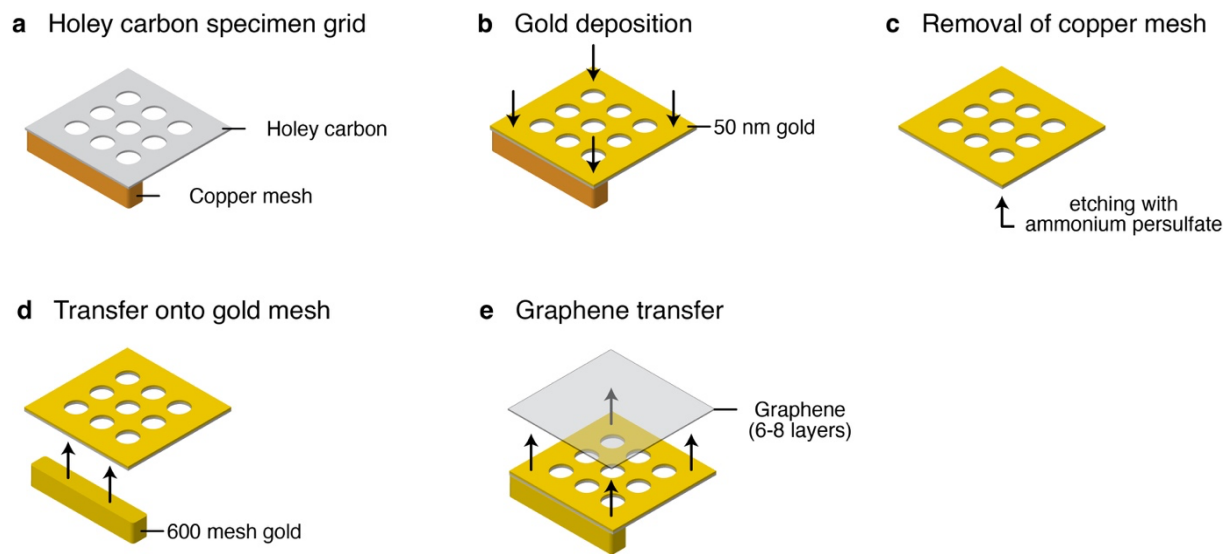

**Figure S2. Fabrication of the sample support. a,b,** A holey gold thin film is fabricated by depositing gold onto a holey carbon film on copper mesh. **c-e,** After etching away the copper, the holey gold film is picked up with a 600 mesh gold grid, and multilayer graphene is transferred onto the assembly.

## 3 Deposition of amorphous solid water

Amorphous solid water is deposited as previously described.(*4*) The sample support is loaded into the microscope with a single tilt cryo specimen holder (Gatan 914), which is then filled with liquid nitrogen to cool the sample to a temperature of 101±1 K, which is continuously monitored during the experiment. Amorphous solid water (ASW) is deposited by leaking water vapor into the column of the microscope through a leak valve that we installed on our instrument for this purpose. Deionized water (Milli-Q 15 MΩ cm) is placed in a stainless-steel reservoir that is water cooled to a temperature of 290 K. After filling the reservoir, the water is degassed by evacuating the reservoir for 5 minutes with a membrane pump. The water vapor in the reservoir is then leaked into the microscope through a gas dosing valve (Balzers UDV235) that is connected to a 60 cm long stainless-steel tube. The end of the tube protrudes into the cold shield, a small metal enclosure between the pole pieces of the objective lens that surrounds the sample, which is ordinarily cooled to liquid nitrogen temperature to serve as a cryo pump, but in our experiments is held at ambient temperature. The nozzle of the tube has an inner diameter of 1 mm and terminates at approximately 10 mm from the edge of the sample holder. Since the nozzle lies in the sample plane, no direct line of sight exists between the tube and the sample. Water molecules must therefore first undergo collisions with surfaces before they can reach the specimen grid. Accordingly, we find that the deposition rate does not vary across the specimen support. The sample region is pumped through openings in the cold shield, with the surrounding volume of the microscope column evacuated to a pressure of $2 \cdot 10^{-7}$ mbar by a 150 L ion pump when no water is being leaked into the microscope.  We adjust the deposition rate, about 60 nm/min, by monitoring the pressure on the low-pressure side of the leak valve.

The following procedure is used to control the thickness of the ASW layer that is freshly deposited for each experiment. Before beginning an experimental run, the deposition rate is determined by recording diffraction patterns as a function of deposition time, which are then analyzed as described in Supporting Note 5. As shown in Fig. S3, the relative intensity of the diffraction pattern initially increases linearly with the deposition time, but then goes through a maximum as the sample grows thicker and multiple scattering as well as inelastic interactions become more important. By comparing this curve to a reference, we can deduce the deposition rate (typically 60 nm/min) and thus adjust the deposition time (typically 4 min) to obtain a sample thickness of 263±20 nm (blue arrow in Fig. S3). Experiments are only started once the deposition rate has stabilized. The deposition rate is remeasured approximately every 2 h, and the deposition time is adjusted accordingly.

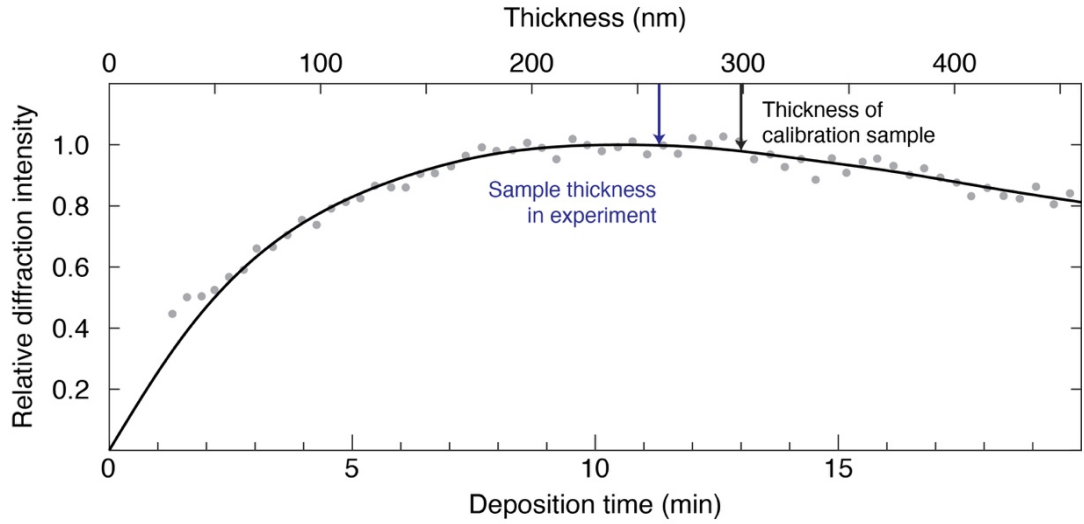

**Figure S3. Relative intensity of the diffraction pattern of water as a function of deposition time.** The corresponding sample thickness is indicated on the top axis, as determined from a calibration experiment. The thickness of the calibration sample (299 nm) as well as the sample thickness in our experiment (263 nm) are marked with black and blue arrows, respectively. The solid line is a spline of the data that serves as a guide to the eye.

The absolute sample thickness was determined with the log-ratio method.(*5, 6*) A calibration sample was grown to a thickness of 299 nm (black arrow in Fig. S3) and transferred to a JEOL 2200FS transmission electron microscope, which is equipped with an energy filter, allowing us to record energy loss spectra of the ASW sample.(*7*) The sample thickness $d$ is obtained from the ratio of the total electron beam intensity $I_t$ to the intensity of the zero loss peak $I_0$,

$$d = \lambda \cdot \ln \frac{I_t}{I_0} \, .$$

Here $\lambda$ and is the electron inelastic mean free path, which can be calculated according to

$$\lambda \approx \frac{106 \, F \, E_0}{E_m \, ln\left(\frac{2\beta E_0}{E_m}\right)},$$

where $F = 0.618$ is a relativistic factor, $E_0$ the accelerating voltage in keV, $\beta = 10$ mrad, and $E_m = 7.12$ the average energy loss in eV.(*6*) The value for $E_m$ was interpolated from recent measurements on thin liquid water films.(*8*) We obtain a thickness of 299 nm for the calibration sample, which yields a thickness of 263 nm for the samples used in our experiments.

**4 Time-resolved diffraction experiments**

The following procedure is adopted to establish suitable experimental parameters. As illustrated in Fig. 1a,b, time-resolved electron diffraction patterns are collected from within the central hole of a grid square, with the electron beam converged to a disk of about 1.5 µm diameter. Care is taken to choose a sample area without any tears in the graphene film. The laser beam is centered onto the same hole in the gold film using the method previously described.(*9*) Once the deposition rate has been established (Supporting Note 3), a 263 nm thick layer of ASW is deposited, and the approximate minimum laser power is determined (approximately 100 mW), with which this layer can be successfully melted and revitrified three times with a 30 µs laser pulse.(*9*) We find that this is only possible in a very narrow range of laser powers of about 10 mW — at higher laser powers, the sample evaporates, while at lower laser powers, it crystallizes. This ensures that the sample undergoes a closely similar temperature evolution in each experiment. The sample is then irradiated with a 1 s laser pulse of approximately 80 mW power to evaporate any ice within the grid square under observation. Finally, the sample is irradiated with another 3 s laser pulse from a second laser (405 nm, 150 mW, beam diameter of approximately 70 µm in the sample plane, centered on the area under observation) in order to also evaporate the ice in adjacent grid squares.

Once the experimental parameters have been established, time-resolved experiments are performed with an automated procedure. A fresh layer of ASW is deposited, and a diffraction pattern is recorded (20 boosted electron pulses of 2 µs duration, fired at 10 Hz repetition rate) which is later used to accurately determine the camera length in each experiment (Supporting Note 5) and normalize the intensity of the time-resolved diffraction pattern. The sample is then melted with a 30 µs laser pulse, and its structural evolution is probed at a given point in time by capturing a diffraction pattern with a boosted electron pulse of 2 µs duration (Supporting Note 1). After the end of the laser pulse, the sample cools rapidly and vitrifies, so that HGW is obtained. The same process is then repeated two more times for the HGW sample — a diffraction pattern is recorded with 20 electron pulses, the sample is melted with a 30 µs laser pulse (same power), and its structural evolution during the melting process is probed with a 2 µs electron pulse. After the experiment, another diffraction pattern is captured (20 electron pulses), which serves to verify that the sample did not evaporate during the last laser pulse, but was successfully melted and revitrified. Finally, the ice within the grid square is evaporated as described above. After every third experiment, the ice in the adjacent grid squares is also evaporated. A fresh layer of ASW is then deposited, and the experiment is repeated. The total time required to melt the sample three times and record all seven diffraction patterns is approximately 65 s.

The structural evolution of ASW and HGW during laser melting was captured five times between 0 µs and 30 µs in single microsecond time steps. The diffraction patterns for each time step were then analyzed as described in Supporting Note 5 and averaged to give the diffraction patterns shown in Fig. 2 b-f.

**5 Analysis of the diffraction patterns**

Both time-resolved and static diffraction patterns are analyzed as previously described.(*4*) We first determine the center of each diffraction pattern from the center of mass of the direct beam and then correct for the ellipticity of the diffraction pattern (typically about 1 %). In each experimental run, the distortion is determined from a diffraction pattern of the polycrystalline holey gold film of the sample support, which we record under identical conditions. The distortion parameters are extracted in analogy to a frequently used method for determining the magnification distortion of electron micrographs from their diffractograms.(*10*) We then azimuthally average the diffraction patterns and subtract the diffraction background, which consists of the atomic scattering term, contributions from inelastic and multiple scattering, as well as the instrument background.(*11*) The diffraction background is determined from a logarithmic spline of the diffraction intensity. For each time-resolved diffraction pattern, the camera length is determined from the static diffraction pattern of the ASW sample that we record before the time-resolved experiment. In order to account for variations in sample thickness as well as the probe current of the laser boosted electron pulses, we normalize the intensity of the time-resolved diffraction pattern on the intensity of this static diffraction pattern, which typically varies by 14 %. This variation is largely due to the fluctuations in the intensity of the boosted electron pulses. The positions and intensities of the first diffraction maximum shown in Fig. 3 and Fig. S4 are determined from 7th order polynomial fits, with the errors calculated from the standard errors of the fit through error propagation.

**6 Determination of the onset of crystallization**

We determine the time at which crystallization sets in with the following procedure. Figure S4 shows the evolution of the intensity of the first diffraction maximum during three successive laser melting experiments as shown in Fig. 3b, with the ASW sample in black and the HGW sample in blue, as well as the HGW sample obtained after melting and revitrifying a third time in green (HGW$_2$). A linear fit of the combined data points of all three experiments at early times (0–6 μs for ASW as well as 0–5 μs for HGW and HGW$_2$) is then subtracted from the transients (red, dashed line), and the curves for ASW and HGW are shifted along the time axis and scaled along the diffraction intensity axis to maximize agreement with the curve for HGW$_2$ (Fig. S4b). This is achieved by comparing the shifted and scaled data points for HGW and ASW to interpolated points on the HGW$_2$ curve and performing a least-squares minimization. The combined data points in the range of 5–7.5 μs are then fit with a straight line (orange, solid line), whose intersection with the horizontal axis (red, dashed line) at 5.33 ± 0.14 μs is taken as the onset of crystallization for the HGW$_2$ sample. The onsets for the ASW (6.79 ± 0.16 μs) and HGW (5.75 ± 0.15 μs) samples, are obtained by adding the time delays determined earlier by which the corresponding transients were shifted to maximize agreement with the HGW$_2$ transient. The errors are calculated by uncertainty analysis and error propagation.

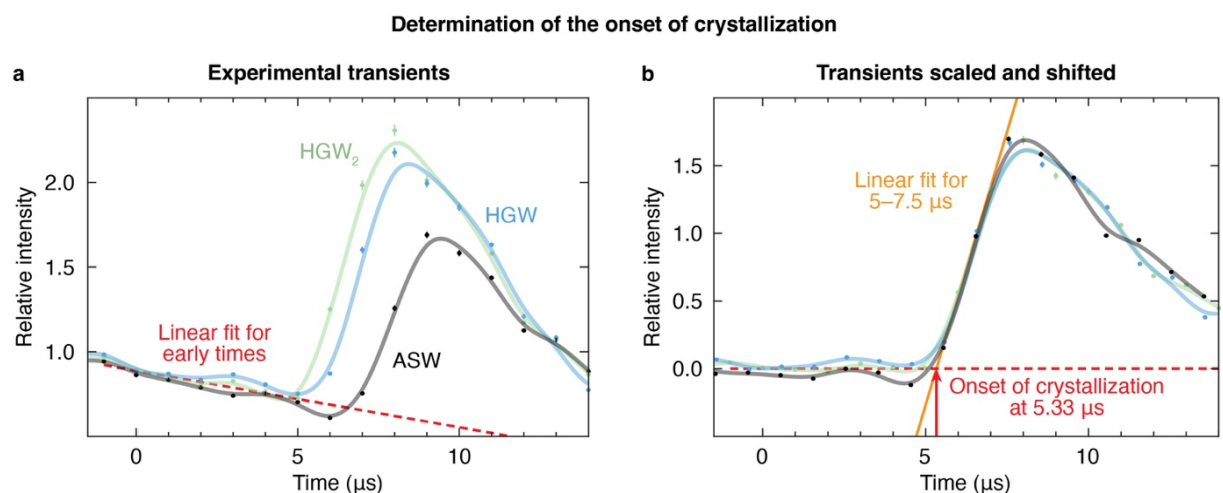

**Figure S4. Illustration of the procedure for determining the onset of crystallization. a,** Temporal evolution of the intensity of the first diffraction maximum during three successive laser melting experiments with ASW in black and HGW in blue, as well as HGW$_2$ in green, as shown in Fig. 3b. Error bars indicate standard errors of the fit of the diffraction intensities. The solid lines provide a guide to the eye and are derived from splines. The dashed line represents a linear fit of the combined data points of all three experiments in the range of 0-6 µs for ASW and 0–5 µs for HGW and HGW$_2$. **b**, The linear fit of the early data points in (**a**) is subtracted from all three curves, and the transients for ASW and HGW are shifted along the time axis and scaled along the intensity axis to maximize agreement with the curve for HGW$_2$. The combined data points in the range of 5–7.5 µs are then fit with a straight line (orange solid line), whose intersection with the horizontal axis (red dashed line) at 5.33 µs is taken as the onset of crystallization for the HGW$_2$ sample. The onsets for the ASW and HGW samples are obtained by adding the time delays by which the corresponding transients were shifted earlier to maximize agreement with the HGW$_2$ transient.

**7 Determination of the sample temperature at the end of the laser pulse**

The temperature of the sample at the end of the laser pulse is determined with the following procedure. The diffraction patterns between 25 µs and 30 µs are averaged, and the position of the first maximum (Supplementary Note 5) is compared to x-ray data (*12*) as described in more detail previously,(*4*) which yields a temperature of 281 ± 2 K. The procedure takes into account that electron and x-ray diffraction show small differences in the positions of the diffraction maxima, which largely arise from differences in the diffraction background that is obtained with either method.(*11*) In order to make it possible to compare the electron and x-ray diffraction data, the momentum transfer axes are adjusted linearly, so that the peak positions of amorphous ice in both experiments coincide (Fig. S10 in Ref. (*4*)). For the x-ray data, we use the peak positions of low density amorphous ice,(*13*) which are compared to the peak positions of HGW in our experiment.

For the experiments of Fig. 1d,e, which were conducted at reduced laser power, the sample temperature at the end of the laser pulse was estimated from heat transfer simulations (Supplementary Note 8), in which the simulated laser power was reduced proportionally. Note that the simulations neglect the heat released due to crystallization.

## 8 Simulation of the temperature evolution of the sample

Finite element heat transfer simulations of the temperature evolution of the sample under illumination with microsecond laser pulses are performed using COMSOL Multiphysics, as previously described.[1](4) The simulation geometry is illustrated in Fig. S5. A 600 mesh gold grid (13.5 μm wide and 8.75 μm thick bars, 27 μm × 27 μm viewing area) supports a thin film of 50 nm gold on 12 nm amorphous carbon that features a square pattern of holes (2 μm diameter, 1 μm separation) and is covered with a film of 7-layer graphene (2.415 nm thickness). The holey thin film and graphene are covered on both sides by a 131.5 nm thick layer of ice, for a total of 263 nm, as shown in Fig. S5b. In simulations of the second and third melting and revitrification experiment, the thickness of the sample near the center of the laser focus is reduced to 235 nm and 204 nm, respectively, to account for evaporation (Supplementary Note 9). The thickness is reduced within a radius of 4 μm, which corresponds approximately to the area that is melted in our experiment. To reduce the computational cost, the graphene film is omitted outside of a 9 μm × 9 μm square area in the center of the geometry, and the extent of the ice-covered holey gold/carbon film is limited to a square of 125 μm side length. To account for the large heat capacity of the specimen grid, the gold bars of the supporting mesh extend another 42.5 μm beyond this square.

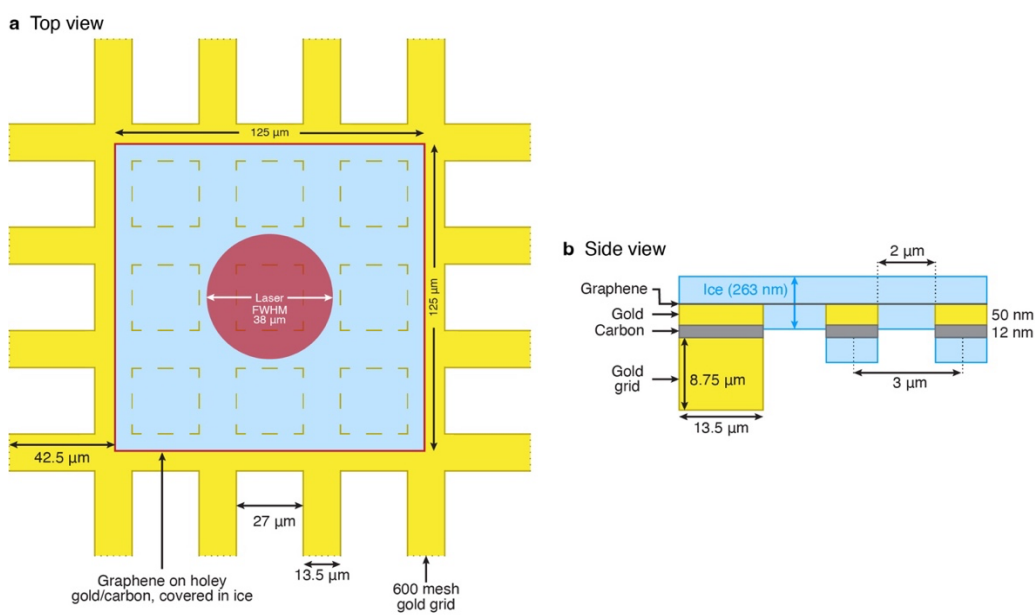

**Figure S5. Sample geometry used in the heat transfer simulations. a**, Top view. The laser is centered on the simulation geometry, with the FWHM of the Gaussian laser spot indicated by a red circle. **b**, Side view.

The material parameters used in our simulations are listed in Table S1. We use literature values for the temperature-dependent heat capacity and thermal conductivity of amorphous carbon,[14] gold,[15, 16] and graphene.[17, 18] Since reliable low-temperature values for the heat capacity of amorphous carbon are not available,[19] we use its room temperature value[20] and extrapolate it to low temperatures by assuming that the temperature dependence is the same as for graphite.[17] Experimental values for the heat capacity and thermal conductivity of water are not available for a wide range of temperatures in no man's land.[12, 21] For the heat capacity of water, we use experimental values between 227 K and 350 K (Refs.[22–24]) as well as between 100 K and 136 K (Ref. [25]) and perform a spline interpolation at intermediate temperatures. For the thermal conductivity of water, we use its room temperature value, which is similar to that of amorphous ice at low temperatures.[26]

**Table S1. Material properties used in the heat transfer simulations.**

| Property | Value | Ref. |
|---|---|---|
| Heat capacity of gold | $38.5679 + 1.2434 \cdot T - 7.137 \cdot 10^{-3} \cdot T^2 + 1.9237 \cdot 10^{-5} \cdot T^3 - 1.9801 \cdot 10^{-8} \cdot T^4$ (J/kg·K) | (15) |
| Thermal conductivity of gold | $320.973 - 0.0111 \cdot T - 2.747 \cdot 10^{-5} \cdot T^2 - 4.048 \cdot 10^{-9} \cdot T^3$ (W/m·K) | (16) |
| Heat capacity of graphene | Data from Table 2 of Ref. (17) | (17) |
| Thermal conductivity of graphene | Data for supported graphene from Fig. 3a of Ref. (18) | (18) |
| Heat capacity of water | Splined data from Ref. (22–25) | (22–25) |
| Thermal conductivity of water | 0.6 W/(m·K) | (26) |
| Absorption of thin film (ice – graphene – gold – carbon – ice) | 37.1 % (263 nm), 38.9 % (235 nm), and 41.4 % (204 nm) Refractive indices used in calculating the absorption at 532 nm ice 1.302, graphene 2.67 + i1.34, gold 0.47 + i2.17, carbon 2.279 + i0.634 | (27–30) |
| Absorption of free-standing graphene (ice – graphene – ice) | 11.7 % (263 nm), 11.0 % (235 nm), and 10.7 % (204 nm) at 532 nm wavelength for 7 layer graphene (2.415 nm thickness) | (27, 28) |
| Enthalpy of evaporation of water | $2.498 \cdot 10^6 - 3.369 \cdot 10^3 \cdot T$ (J/kg) | (31) |
| Evaporation rate | $\frac{\gamma_e P_0}{\sqrt{2\pi m k_B T}}$, where $\gamma_e = 1.0$ and $P_0$ is the vapor pressure as defined in Eq. 2 of Ref. (32) | (32) |
| Vapor pressure of water | Equation 1 on page 350 of Ref. (33) | (33) |

The temperature of the entire sample is initially set to 101 K. We simulate heating with a 30 µs laser pulse by placing two Gaussian surface heat sources (38 µm FWHM) in the center of the simulation geometry. In order to account for the different absorption of the graphene covered gold film and of the free-standing graphene areas, one heat source is placed on top of the gold film and one on top of the free-standing graphene. The fraction of the laser intensity absorbed by the sample depends on the thickness of the ice layer and was determined with the help of a multilayer thin film transfer matrix calculator,(34) using the complex refractive indices of amorphous ice,(27) graphene,(28) gold,(29) and amorphous carbon.(*30*) The heating rates of the Gaussian surface heat sources are calculated from the absorbed fractions and the incident laser power. The temporal profile of the simulated laser pulse (rectangular pulse with rise and fall times of 250 ns, which we model with error functions) closely mimics that in the experiment. The laser power of 223 mW is chosen such that the sample reaches a temperature of 281 K at the end of the laser, as determined experimentally (Supplementary Note 7). To account for evaporative cooling, we place negative heat sources on the top and on the bottom surface of the water film. The cooling rate is determined from the temperature-dependent enthalpy of evaporation (*31*) and the temperature-dependent evaporation rate of water, as calculated with the formula in Table S1.(*32*) The temperature of the ice is probed within the hole of the gold film that is located in the center of the geometry. We report the average temperature during the electron pulse within a cylindrical volume of 1.5 μm diameter that spans the entire thickness of the ice film and that is centered on the central hole.

**9 Simulation of the crystallization kinetics**

We simulate the crystallization kinetics during the three successive laser melting experiments in order to verify that the slower crystallization of ASW cannot be trivially explained, but instead points to different crystallization kinetics of supercooled liquids that are formed upon laser melting ASW and HGW. Performing an accurate simulation encounters two difficulties. We do not know precisely how much the sample thins due to evaporation with each laser pulse. Moreover, the nucleation rate of water in no man's land is largely unknown.(*35*) We therefore use the following iterative procedure. We first simulated the temperature evolution of the sample in each laser melting experiment, based on a guess of how much the sample thins with each laser pulse. Using the simulated temperature evolution thus obtained, we then simulate the crystallization kinetics for HGW in the second laser melting experiment and adjust the nucleation rate until the onset of crystallization occurs at the same time as in our experiment (5.75 µs, Fig. S4). With this adjusted nucleation rate, we then simulate the crystallization kinetics for all three laser melting experiments and determine the time delay between the onset of crystallization during the second and third experiment (HGW and HGW$_2$ in the nomenclature of Fig. 3). We then update our guess of how much material is evaporated with each laser pulse and iterate until this time delay matches our experimental observation (0.42 µs, Fig. S4). The steps of this iterative process are detailed in the following.

Determination of the sample thickness and simulation of the temperature evolution

We initially simulate the temperature evolution of the sample for a range of thicknesses between 263 nm, the initial sample thickness, and 190 nm (Fig. S6).

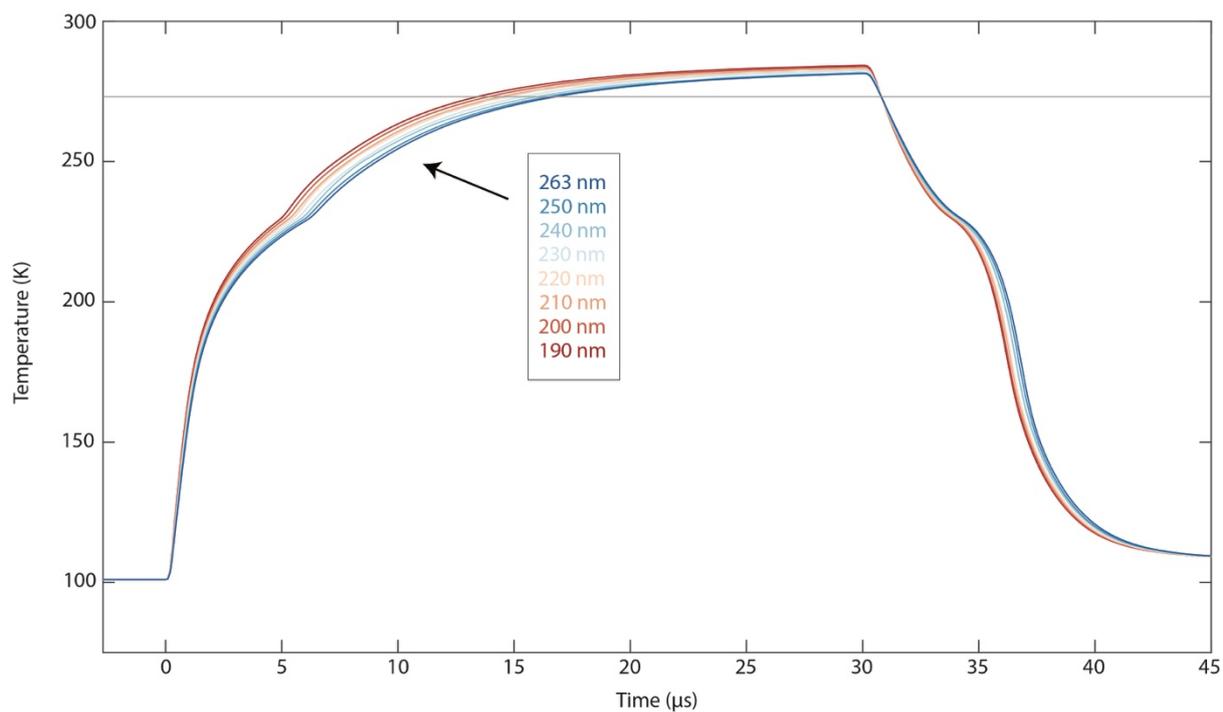

**Fig. S6. Simulated temperature evolution of the sample for thicknesses between 263 nm and 190 nm.**

In each of the three laser melting experiments part of the sample evaporates in the vacuum of the electron microscope. Figure S7 shows that the amount of material evaporated per laser pulse increases approximately linearly with decreasing initial sample thickness, since thinner samples heat up more rapidly, so that more material is evaporated. Here we have calculated the evaporated thickness by integrating the temperature dependent evaporation rate (Table S1) over the simulated temperature evolution of the samples of the different thicknesses shown in Fig. S6.

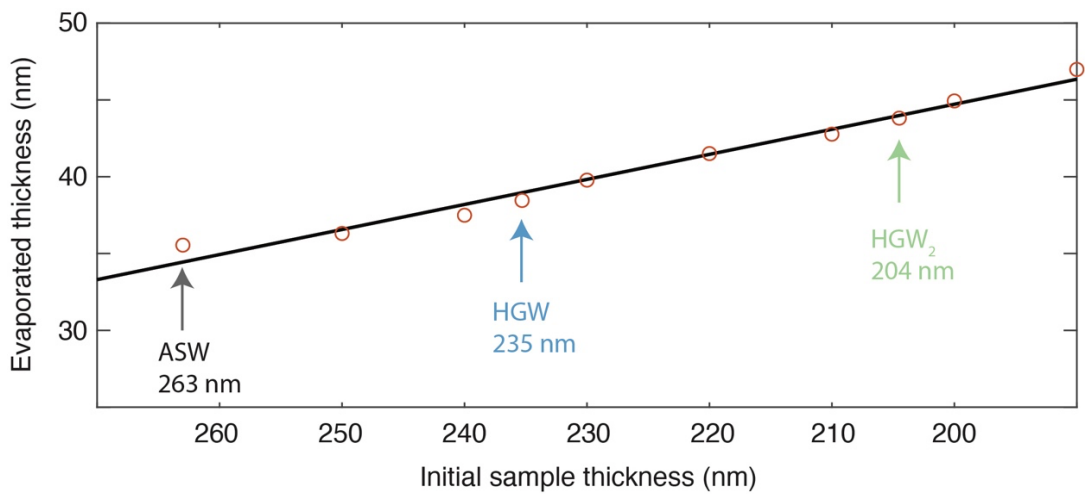

**Figure S7. Sample thickness that is evaporated during the laser pulse as a function of the initial sample thickness, as determined from simulations of the temperature evolution in Fig. S6.** The black line represents a linear fit.

The sample thickness decreases by less than the amount that is evaporated by each laser pulses since some amount of ASW is deposited between laser melting experiments — about 15 nm, as estimated from the deposition rate (Supplementary Note 4). In order to account for the uncertainty in the deposition rate, we treat the amount of ASW deposited as a free parameter, which simultaneously serves as a zero-order correction for any other errors in our determination of the evaporated thickness. From the thickness evaporated in the first laser pulse (36 nm) and the amount of ASW deposited (initial guess of 15 nm) we can calculate the sample thickness in the second experiment. We then obtain the temperature evolution for a sample of this thickness by interpolating the simulated data in Fig. S6. We again calculate the thickness that is evaporated during this second laser pulse to obtain the sample thickness in the third experiment, for which we obtain the temperature evolution through interpolation of the simulated data in Fig. S6 as above.

As detailed below, the simulations of the crystallization kinetics reproduce our experimental observations if 8 nm of ASW are deposited between laser pulses, with the sample thickness reduced from 263 nm in the first melting experiment to 235 nm and 204 nm in the second and third experiment, respectively. Figure S8 shows the corresponding simulations of the temperature evolution. At 5.75 µs, when the HGW sample begins to crystallize in our experiment, it reaches a temperature of 229 K. At this temperature, the time delay between the first and second laser melting experiment is 0.402 µs (Figure S8, inset). The time delay between the second and third experiment is somewhat larger (0.540 µs) due to the increased evaporation.

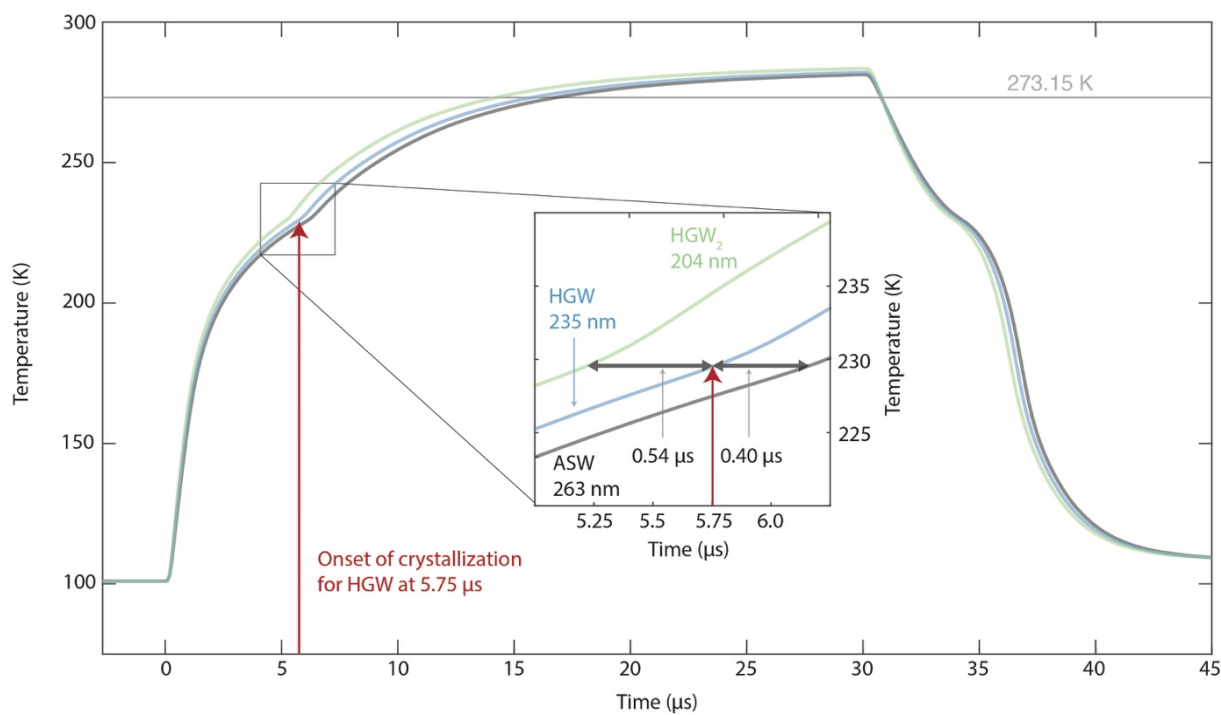

**Figure S8. Simulated temperature evolution of the sample during three successive laser melting experiments.** Evaporation decreases the sample thickness from 263 nm (ASW, black) to 235 nm (HGW, blue), and 204 nm (HGW$_2$, green), which causes the sample to heat up more rapidly and cool down more rapidly after the end of the laser pulse.

Note that our simulations neglect that the heat released by the crystallization process causes the sample to warm up more rapidly. We estimate that this leads to the evaporation of an additional 5 nm per laser pulse. However, since the additional evaporated thickness is nearly identical in all three laser melting experiments (changing by less than 1 nm), we do not need to explicitly consider it.

<u>Simulation of the crystallization kinetics</u>

Once we have determined the temperature evolution of the sample during each laser pulse, we simulate the crystallization kinetics with a simple model based on classical nucleation and growth theory.(*36*, *37*) It assumes that nuclei form with the nucleation rate $J(T)$, after which they grow into spherical crystals with the growth rate $\rho(T)$. Figure S9a shows the nucleation rate $J(T)$ used in our simulation (solid curve), which we obtain from a fit of experimental data (*35*) (circles) with an asymmetric super-Gaussian.

$$\ln\left(J(T)\right) = \ln\left(J_{\max}\right) \cdot e^{-\left(\frac{T-T_0+s*(T-T_0)^2}{c}\right)^4}$$

Where $J_{max}$ is the maximum nucleation rate, $T_0$ the center, $c$ the width, and $s$ the skewing factor of the asymmetric super-Gaussian.

The nucleation rates in no man's land are largely unknown.(*35*) We therefore adjust $J_{max}$, which in our fit corresponds to the plateau of the super-Gaussian, such that our simulation reproduces the onset of crystallization for HGW in the second laser melting experiment at 5.75 µs (Fig. S4a). This yields a value of $J_{max}$ = 3.4·10²⁶ m⁻³s⁻¹ somewhat above the upper bound of a previous estimate of the nucleation rates, which was obtained from hyperquenching experiments on microdroplets (horizontal lines with $J_{max}^{microdroplets}$ = 1.5·10²² m⁻³s⁻¹ and $J_{min}^{microdroplets}$ = 1.5·10¹⁹ m⁻³s⁻¹.(*35*) The growth rate (Fig. S9b) is obtained from a logarithmic spline of experimental data.(*38*)

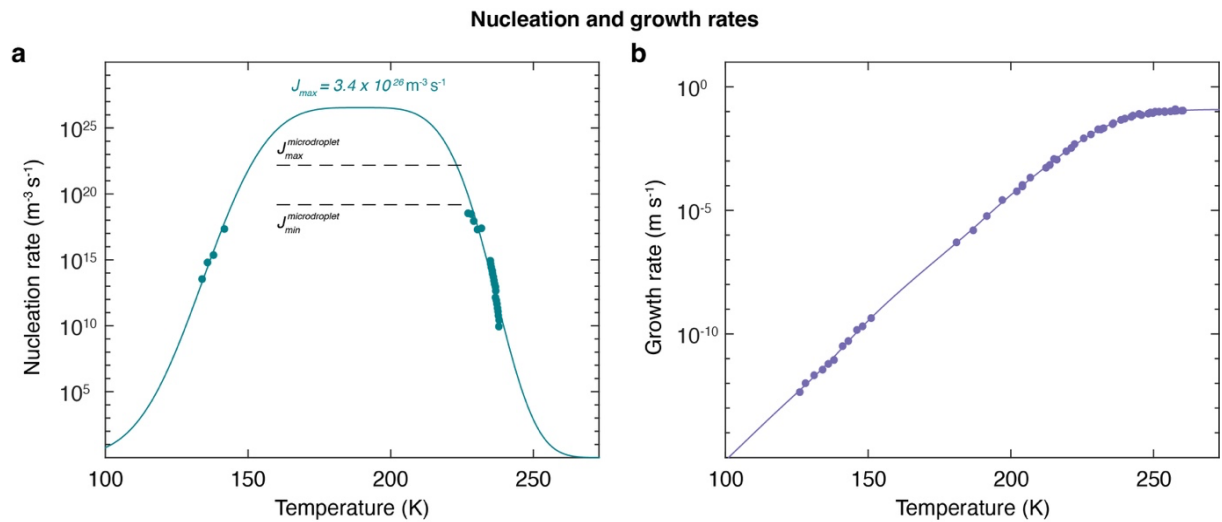

**Figure S9. Nucleation and growth rates used in the simulations. a**, The nucleation rate is obtained by fitting an asymmetric super-Gaussian (solid curve) to experimental data (dots),(*35*) with the maximum nucleation rate set to $J_{max}$ = 3.4·10²⁶ m⁻³s⁻¹, so that our simulation reproduces the crystallization kinetics of HGW. **b**, The growth rate is obtained from a logarithmic spline of experimental data.(*38*)

As long as the crystalline fraction $f(t)$ is small, it can be calculated as

$$f(t) = \int_0^t J(T(t')) \frac{4\pi}{3} \left( \int_{t'}^t \rho(T(t'')) \, dt'' \right)^3 dt'$$

where $\frac{4\pi}{3} \left( \int_{t'}^t \rho(T(t'')) \, dt'' \right)^3$ is the volume of a spherical crystal at time $t$ that was nucleated at time $t'$ and $T(t)$ is the temperature evolution of the sample, as obtained from our heat transfer simulations. Both nucleation and growth slow once a large fraction of the sample $f(t)$ has already crystallized. We account for this effect by scaling both nucleation and growth rates with a factor $[1 - f(t)]$.

$$f(t) = \int_0^t [1 - f(t')] J(T(t')) \frac{4\pi}{3} \left( \int_{t'}^t [1 - f(t'')] \rho(T(t'')) \, dt'' \right)^3 dt'$$

Even though this approximation is somewhat simplistic, this is of little consequence for our analysis, which uses the simulations only to determine the time at which crystallization sets in (*i.e.* $f(t)$ is small), which we then compare to the experiment. For the same reason, we can also neglect that once the sample starts crystallizing, it heats up more rapidly than in our simulation, which does not take into account the heat released by the crystallization process.

We determine the crystalline fraction $f(t)$ through numerical integration, starting with an initial guess of $f(t) = 0$ and iterating until $f(t)$ has converged. For ASW, we start the simulation at the beginning of the laser pulse ($t = 0$ µs), whereas for the HGW samples, we also include the preceding hyperquenching process that follows after the end of the previous laser pulse. Since the HGW (HGW$_2$) sample traverses the supercooled regime twice, the final concentration of nuclei is about 1.49 times (1.40 times) as large as for ASW, which traverses it only once. As discussed above, we adjust the maximum nucleation rate in no man's land to $J_{max}$ = 3.4·10$^{26}$ m$^{-3}$s$^{-1}$, so that in the simulation of the second laser melting experiment, crystallization sets in at 5.75 µs, as observed experimentally (Fig. S4a). Here, we have chosen to set the crystalline fraction at this point in time to 0.01. This is a somewhat arbitrary choice. However, if we instead reduce the crystalline fraction to 0.001, our result changes only nominally, with the time delay in the onset of crystallization between the first and second laser melting experiment changing by only 3 ns (while the maximum nucleation rate $J_{max}$ increases by about one order of magnitude).

Figure S10 shows the simulated crystalline fraction as a function of time. At the onset of crystallization ($f(t) = 0.01$), the time delay between the first and second laser melting experiment is 0.48 μs (inset). This is somewhat longer than the time delay of 0.40 μs that results from the change in thickness (Fig. S8) and reflects the fact that the ASW sample in the first experiment traverses no man's land only once and therefore forms fewer nuclei. The time delay between the second and third experiment is somewhat smaller (0.42 μs) than expected merely based on the change in sample thickness (0.54 μs, Fig. S8). This is consequence of the fact that the sample in the third experiment is thinner and thus heats up and cools down faster, so that it spends less time in no man's land and forms fewer nuclei.

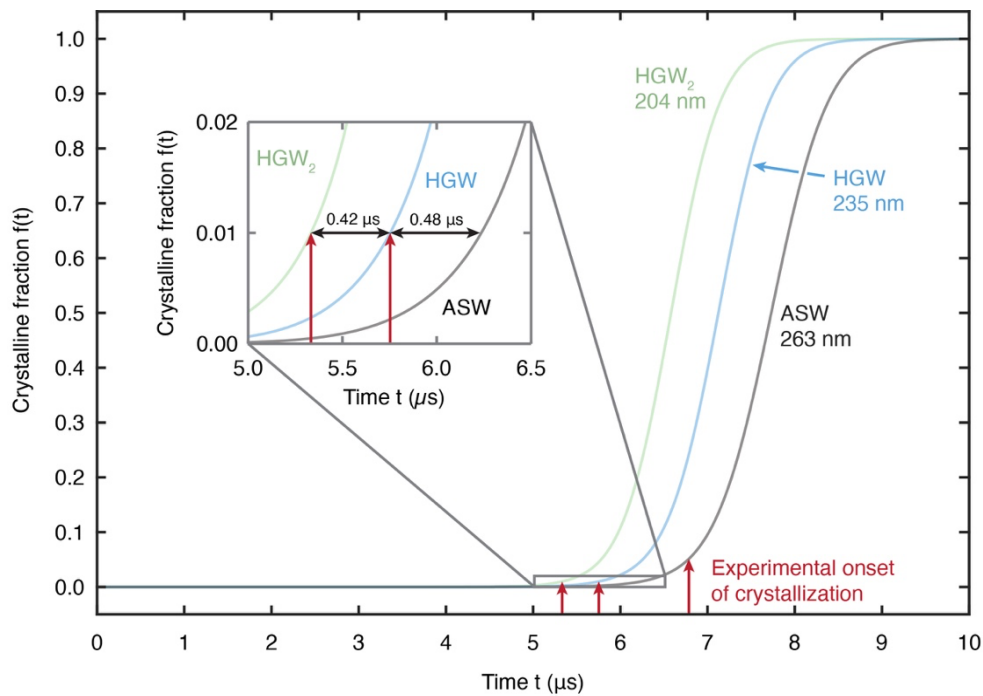

**Figure S10. Simulation of the crystallization kinetics of the sample during three successive laser melting experiments.** Simulated crystalline fraction as a function of time for the first experiment (ASW, black, 263 nm sample thickness), the second (HGW, blue, 235 nm), and the third experiment (HGW₂, green, 204 nm). In the inset, the time delays between the three experiments are indicated for the onset of crystallization (crystalline fraction 0.01).

Note that after the onset of crystallization, the crystalline fraction grows faster in our simulations than we observe experimentally. For example, we estimate that in our experiment, only about 30 % of the sample have crystallized when it reaches the melting point at about 8.5 µs (Supplementary Note 10), while the simulation predicts that the sample has fully crystallized at that point. Evidently, our simulations overestimate the rate of crystal growth. This is probably due to the fact that nucleation likely occurs at the graphene and vacuum interfaces of the sample(*37*) and that because of geometric constraints, crystals at the interface grow more slowly. Note that this discrepancy between simulation and experiment does not affect our analysis, in which we only consider the onset of crystallization.

In conclusion, our simulations predict that in each subsequent laser melting experiment, crystallization sets in earlier by about the same time delay, 0.48 µs between the first and second experiment and 0.42 µs between the second and third. This is largely explained by the thinning of the sample due to evaporation during the laser pulse and the resulting speed-up of the laser heating process. In contrast, we find in our experiments that the ASW sample crystallizes about 1.04 µs later than the HGW sample. This evidently cannot be trivially explained by changes in the sample thickness. Instead, it suggests that the supercooled liquids that are formed upon melting either sample have different crystallization kinetics. The simulations reproduce the onset of crystallization observed for ASW if the maximum nucleation rate is lowered by a factor of 5 to $J_{max}$ = 6.6·$10^{25}$ m$^{-3}$s$^{-1}$ or alternatively, if we decrease the growth rate by a factor of 1.7 at all temperatures.

## 10 Estimate of the maximum crystalline fraction

We estimate the maximum crystalline fraction of the sample during laser melting from the fact that in our experiments, the samples reach the melting point sooner than in our simulation due to the heat released by the crystallization process. We determine that the ASW, HGW, and HGW$_2$ sample reach the melting point at 9.5 µs, 8.5 µs, and 8.0 µs, respectively, when the intensity of the first diffraction peak reaches a maximum (Fig. 3a). At these times, our heat transfer simulations predict a sample temperature of $T_o$ = 252.1 K, 250.0 K, and 251.6 K, respectively. The increase in temperature in our experiment must therefore be due to the heat released by the crystallization process.

The latent heat of melting $H_{melting}(T_o)$ at the temperature $T_o$ can be determined using Kirchhoff's relation(*39*)

$$H_{melting}(T_o) = \int_{T_o}^{T_m} C_{p,ice} \, dT + H_{latent}(T_m) - \int_{T_o}^{T_m} C_{p,water} \, dT,$$

where $T_m$ is the melting point of water, $H_{latent}(T_m)$ is the latent heat of melting at the melting point, and $C_{p,ice}$ and $C_{p,water}$ the heat capacities of ice and water, respectively (Table S1 and Supporting Note 8). The energy $\Delta H(T_0)$ required to heat the sample from the the temperature $T_o$ to the melting point $T_m$ is

$$\Delta H(T_0) = (1-f) \int_{T_o}^{T_m} C_{p,water} \, dT + f \int_{T_o}^{T_m} C_{p,ice} \, dT,$$

where $f$ is the crystalline fraction to be determined. At the same time, the following relationship for the crystalline fraction $f$ must hold

$$f = \frac{\Delta H(T_0)}{H_{melting}(T_o)}.$$

By solving for the crystalline fraction $f$, we can then estimate that about 30 % of the sample crystallizes.

From the relative maximum intensity of the first diffraction maximum, we can furthermore estimate that the maximum crystalline fraction of the HGW and $HGW_2$ samples is about 1.5 times as high as for ASW, where we have used the procedure detailed in Supporting Note 6 for our estimate.

## 11 Formation of different ice polymorphs during laser melting

Crystallization during laser melting initially involves the formation of stacking disordered ice, which transforms into hexagonal ice once the sample reaches higher temperatures. This can be seen when comparing time-resolved diffraction patterns of ASW and HGW recorded during laser melting at 8 μs, which show the signature of stacking disordered ice (Fig. S11a), and at 16 μs, where an additional feature at 2.35 $Å^{-1}$ appears that is characteristic of the formation of hexagonal ice (Fig. S11b, feature marked with an arrow).

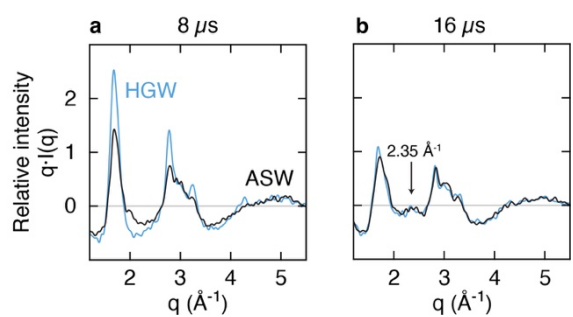

**Figure S11. Time-resolved diffraction patterns of ASW and HGW during laser melting at 8 μs and 16 μs. a**, Time-resolved diffraction patterns recorded at 8 μs show that stacking disordered ice is initially formed. **b**, In contrast, diffraction patterns recorded at 16 μs exhibit the signature of hexagonal ice (characteristic feature marked with an arrow, see also Fig. 1e).

**12 Diffraction peak positions indicated with blue lines in Fig. 1c-f**

The positions of the diffraction maxima of ASW and HGW indicated in Fig. 1c,f were determined from the diffraction patterns in Fig. 1c,f with the procedure described in Supplementary Note 5. The peak positions indicated for stacking disordered ice Fig. 1d were simulated with SingleCrystal™(*40*) an ice structure (50% cubic and 50% hexagonal) generated with GenIce2,(*41*) while the peak positions for hexagonal ice in Fig. 1e were simulated from a 100 % hexagonal structure.

## References

1. G. Bongiovanni, P. K. Olshin, M. Drabbels, U. J. Lorenz, Intense microsecond electron pulses from a Schottky emitter. *Appl. Phys. Lett.* **116**, 234103 (2020).

2. P. K. Olshin, G. Bongiovanni, M. Drabbels, U. J. Lorenz, Atomic-Resolution Imaging of Fast Nanoscale Dynamics with Bright Microsecond Electron Pulses. *Nano Lett.* **21**, 612–618 (2021).

3. M. G. Pamato, I. G. Wood, D. P. Dobson, S. A. Hunt, L. Vočadlo, The thermal expansion of gold: point defect concentrations and pre-melting in a face-centred cubic metal. *J. Appl. Crystallogr.* **51**, 470–480 (2018).

4. C. R. Krüger, N. J. Mowry, G. Bongiovanni, M. Drabbels, U. J. Lorenz, Electron diffraction of deeply supercooled water in no man's land. *Nat. Commun.* **14**, 2812 (2023).

5. S. Keskin, P. Kunnas, N. de Jonge, Liquid-Phase Electron Microscopy with Controllable Liquid Thickness. *Nano Lett.* **19**, 4608–4613 (2019).

6. Malis, T.; Cheng, S. C.; Egerton, R. F., EELS Log-Ratio Technique for Specimen-Thickness Measurement in the TEM. *J Electron Microsc Tech* **8**, 193–200 (1988).

7. P. K. Olshin, M. Drabbels, U. J. Lorenz, Characterization of a time-resolved electron microscope with a Schottky field emission gun. *Struct. Dyn.* **7**, 054304 (2020).

8. M. N. Yesibolati, S. Laganá, S. Kadkhodazadeh, E. K. Mikkelsen, H. Sun, T. Kasama, O. Hansen, N. J. Zaluzec, K. Mølhave, Electron inelastic mean free path in water. *Nanoscale* **12**, 20649–20657 (2020).

9. J. M. Voss, O. F. Harder, P. K. Olshin, M. Drabbels, U. J. Lorenz, Microsecond melting and revitrification of cryo samples. *Struct. Dyn.* **8**, 054302 (2021).

10. T. Grant, N. Grigorieff, Automatic estimation and correction of anisotropic magnification distortion in electron microscopes. *J. Struct. Biol.* **192**, 204–208 (2015).

11. B. E. Warren, *X-Ray Diffraction* (Dover Publications, 1990).

12. J. A. Sellberg, C. Huang, T. A. McQueen, N. D. Loh, H. Laksmono, D. Schlesinger, R. G. Sierra, D. Nordlund, C. Y. Hampton, D. Starodub, D. P. DePonte, M. Beye, C. Chen, A. V. Martin, A. Barty, K. T. Wikfeldt, T. M. Weiss, C. Caronna, J. Feldkamp, L. B. Skinner, M. M. Seibert, M. Messerschmidt, G. J. Williams, S. Boutet, L. G. M. Pettersson, M. J. Bogan, A. Nilsson, Ultrafast X-ray probing of water structure below the homogeneous ice nucleation temperature. *Nature* **510**, 381–384 (2014).

13. D. T. Bowron, J. L. Finney, A. Hallbrucker, I. Kohl, T. Loerting, E. Mayer, A. K. Soper, The local and intermediate range structures of the five amorphous ices at 80K and ambient pressure: A Faber-Ziman and Bhatia-Thornton analysis. *J. Chem. Phys.* **125**, 194502 (2006).

14. A. J. Bullen, K. E. O'Hara, D. G. Cahill, O. Monteiro, A. von Keudell, Thermal conductivity of amorphous carbon thin films. *J. Appl. Phys.* **88**, 6317–6320 (2000).

15. J. W. Arblaster, Thermodynamic properties of gold. *J Phase Equilib Diffus* **37**, 229–245 (2016).

16. J. Huang, Y. Zhang, J. K. Chen, Ultrafast solid–liquid–vapor phase change of a gold film induced by pico- to femtosecond lasers. *Appl. Phys. A* **95**, 643–653 (2009).

17. W. DeSorbo, W. W. Tyler, The specific heat of graphite from 13° to 300°K. *J. Chem. Phys.* **21**, 1660–1663 (1953).

18. E. Pop, V. Varshney, A. K. Roy, Thermal properties of graphene: Fundamentals and applications. *MRS Bull.* **37**, 1273–1281 (2012).

19. *S.R.P. Silva, Properties of Amorphous Carbon, INSPEC, The Institution of Electrical Engineers, London, 2003.*

20. C. Moelle, M. Werner, F. Szücs, D. Wittorf, M. Sellschopp, J. von Borany, H.-J. Fecht, C. Johnston, Specific heat of single-, poly- and nanocrystalline diamond. *Diam. Relat. Mater.* **7**, 499–503 (1998).

21. K. Murata, H. Tanaka, Liquid–liquid transition without macroscopic phase separation in a water–glycerol mixture. *Nat. Mater* **11**, 436–443 (2012).

22. C. A. Angell, W. J. Sichina, M. Oguni, Heat capacity of water at extremes of supercooling and superheating. *J. Phys. Chem.* **86**, 998–1002 (1982).

23. H. Pathak, A. Späh, N. Esmaeildoost, J. A. Sellberg, K. H. Kim, F. Perakis, K. Amann-Winkel, M. Ladd-Parada, J. Koliyadu, T. J. Lane, C. Yang, H. T. Lemke, A. R. Oggenfuss, P. J. M. Johnson, Y. Deng, S. Zerdane, R. Mankowsky, P. Beaud, A. Nilsson, Enhancement and maximum in the isobaric specific-heat capacity measurements of deeply supercooled water using ultrafast calorimetry. *Proc. Natl. Acad. Sci.* **118**, e2018379118 (2021).

24. Eric W. Lemmon, Ian H. Bell, Marcia L. Huber, and Mark O. McLinden, "Thermophysical Properties of Fluid Systems" in *NIST Chemistry WeBook, NIST Standard Reference Database Number 69* (National Institute of Standards and Technology, Gaithersburg, MD).

25. L. M. Shulman, The heat capacity of water ice in interstellar or interplanetaryconditions. *Astron. Astrophys.* **416**, 187–190 (2004).

26. O. Andersson, H. Suga, Thermal conductivity of low-density amorphous ice. *Solid State Commun* **91**, 985–988 (1994).

27. V. Kofman, J. He, I. Loes ten Kate, H. Linnartz, The Refractive Index of Amorphous and Crystalline Water Ice in the UV–vis. *Astrophys. J.* **875**, 131 (2019).

28. X. Wang, Y. P. Chen, D. D. Nolte, Strong anomalous optical dispersion of graphene: complex refractive index measured by Picometrology. *Opt. Express* **16**, 22105 (2008).

29. G. Rosenblatt, B. Simkhovich, G. Bartal, M. Orenstein, Nonmodal Plasmonics: Controlling the Forced Optical Response of Nanostructures. *Phys. Rev. X* **10**, 011071 (2020).

30. W. W. Duley, Refractive indices for amorphous carbon. *Astrophys. J.* **287**, 694 (1984).

31. M. Fuchs, E. Dayan, E. Presnov, Evaporative cooling of a ventilated greenhouse rose crop. *Agric. For. Meteorol.* **138**, 203–215 (2006).

32. J. D. Smith, C. D. Cappa, W. S. Drisdell, R. C. Cohen, R. J. Saykally, Raman thermometry measurements of free evaporation from liquid water droplets. *J Am Chem Soc* **128**, 12892–12898 (2006).

33. R. J. List, *Smithsonian Meteorological Tables* (Smithsonian Institution Press, Washington, 1968).

34. S. J. Byrnes, Multilayer optical calculations. *ArXiv160302720 Phys.* (2020).

35. P. Gallo, K. Amann-Winkel, C. A. Angell, M. A. Anisimov, F. Caupin, C. Chakravarty, E. Lascaris, T. Loerting, A. Z. Panagiotopoulos, J. Russo, J. A. Sellberg, H. E. Stanley, H. Tanaka, C. Vega, L. Xu, L. G. M. Pettersson, Water: A Tale of Two Liquids. *Chem. Rev.* **116**, 7463–7500 (2016).

36.  P. G. Debenedetti, *Metastable Liquids: Concepts and Principles* (Princeton University Press, Princeton, 1997).

37.  E. H. G. Backus, M. L. Grecea, A. W. Kleyn, M. Bonn, Surface Crystallization of Amorphous Solid Water. *Phys. Rev. Lett.* **92**, 236101 (2004).

38.  Y. Xu, N. G. Petrik, R. S. Smith, B. D. Kay, G. A. Kimmel, *Growth Rate of Crystalline Ice and the Diffusivity of Supercooled Water from 126 to 262 K* (Proc Natl Acad Sci USA, 2016)vol. 113.

39.  W. Cantrell, A. Kostinski, A. Szedlak, A. Johnson, Heat of Freezing for Supercooled Water: Measurements at Atmospheric Pressure. *J. Phys. Chem. A* **115**, 5729–5734 (2011).

40.  Generated using SingleCrystal™: a single-crystal diffraction program for Mac and Windows. CrystalMaker Software Ltd, Oxford, England (www.crystalmaker.com).

41.  M. Matsumoto, T. Yagasaki, H. Tanaka, GenIce: Hydrogen-Disordered Ice Generator. *J. Comput. Chem.* **39**, 61–64 (2018).

42.  L. Kringle, W. A. Thornley, B. D. Kay, G. A. Kimmel, Structural relaxation and crystallization in supercooled water from 170 to 260 K. *Proc. Natl. Acad. Sci.* **118**, e2022884118 (2021).